# Quasi-Large Sparse-Sequence CDMA: Approach to Single-User Bound by Linearly-Complex LAS Detectors

Yi Sun
Department of Electrical Engineering
The City College of City University of New York
New York, NY 10031
E-mail: ysun@ee.ccny.cuny.edu

*Abstract* – **We have proposed a quasi-large random-sequence (QLRS) CDMA where $K$ users access a point through a common channel with spectral spreading factor $N$. Each bit is extended by a temporal spreading factor $B$ and hopped on a $BN$-chip random sequence that is spread in time and frequency. Each user multiplexes and transmits $B$ extended bits and the total channel load is $\alpha = K/N$ bits/s/Hz. The linearly-complex LAS detectors detect the transmitted bits. We have obtained that as $B \to \infty$, if $\alpha < \frac{1}{2} - 1/(4\ln 2)$, each transmitted bit achieves the single-bit bound in BER in high SNR regime as if there was no interference bit. In simulation, when bit number $BK \geq 500$, each bit can approach the single-bit bound for $\alpha$ as high as 1 bit/s/Hz. In this paper, we further propose the quasi-large sparse-sequence (QLSS) CDMA by replacing the dense sequence in QLRS-CDMA with sparse sequence. Simulation results show that when the nonzero chips are as few as 16, the BER is already near that of QLRS-CDMA while the complexity is significantly reduced due to sequence sparsity.**

## I. INTRODUCTION

It is has been known that there exist CDMA channels [1][2] where the jointly optimum global maximum likelihood (GML) detector can achieve the single-user performance in high SNR regime. However, the GML with exhaustive search is NP-hard to obtain and thus is impractical.

Inspired by success of low density parity check (LDPC) codes, large sparsely spread (LSS) CDMA emerged recently [9]-[11] where sequences have only a small number $L$ of nonzero chips. The belief propagation (BP) algorithm and its variations can achieve near optimum performance with a low complexity due to sequence sparsity. Its large system performance is analyzed in [9][10]. The key to practically use BP is to have a small number of nonzero chips $L$, whose decrease degrades LSS-CDMA performance. When $L = 16$, the performance is already near that of densely spread CDMA [10]. The BP is exponentially complex and thus the user number cannot be too large either. Guo and Wang [11] demonstrated that when the user number is about one hundred, the performance is near theoretical result of large system limit. Nevertheless, for the BP to approach the GML [9]-[11], the CDMA system must be impractically large both in user number and spectral spreading factor.

To approach the single-user bound with a practical complexity, we recently proposed the quasi-large random-sequence (QLRS) CDMA [7][8]. When channel load is less than $\frac{1}{2} - 1/(4\ln 2)$ bits/s/Hz and noise power is vanishing, a local maximum likelihood (LML) point is almost surely the GML point. The asymptotic multiuser efficiency (AME) of the LML detectors [5][6] converges almost surely to one. Simulation results verify the analytical results in a much broader region of channel load as high as 1.0 and of any SNR with bit number $\geq 500$. Hence, QLRS-CDMA has approached the bound that a perfect TDMA or FDMA can approach.

In this paper, we further propose the quasi-large sparse-sequence CDMA by replacing the dense sequence in the QLRS-CDMA with the sparse sequence. As the number of nonzero chips $L$ decreases, the sequence sparsity can reduce the complexity but may degrade performance. However, simulation results show that when $L = 16$, the BER can be very close to that of dense sequence while complexity is significantly reduced.

## II. PROPOSED QLSS-CDMA

### A. LSS-CDMA

Consider a $K$-user bit-synchronous Gaussian CDMA channel. The bit period is $T_b$ and the chip period is $T_c$. Then the transmission rate for uncoded bits per user is $1/T_b$ bits/s, the channel bandwidth equals approximately $W = 1/T_c$ and the spectral spreading factor $N = T_b/T_c$. The chip matched filter (MF) at the receiver outputs an $N$-dimensional vector

$$\mathbf{r} = \sum_{k=1}^{K} A_k \mathbf{s}_k b_k + \mathbf{m} = \mathbf{SAb} + \mathbf{m}. \quad (1)$$

$\mathbf{A} = \text{diag}(A_1, \ldots, A_K)$ where $A_k$'s are signal amplitudes and $\mathbf{b} = (b_1, \ldots, b_K)^T$ is the vector of transmitted bits of $K$ users. $b_k$ takes on $\pm 1$'s with equal probability. $\mathbf{m} \sim N(\mathbf{0}, \sigma^2 \mathbf{I})$ is a white Gaussian noise vector. $\mathbf{S} = (\mathbf{s}_1, \ldots, \mathbf{s}_K)$ is the matrix of $N$-chip spreading sequences.

Each sequence $\mathbf{s}_k \in \{-1/\sqrt{L}, 0, 1/\sqrt{L}\}^N$ has $L$ nonzero chips that are randomly located and equiprobably take on $\pm 1/\sqrt{L}$'s. A more general distribution of the nonzero chips can be considered, though [10]. Note that each sequence is normalized to unit length $\|\mathbf{s}_k\| = 1$. After spreading, the energy of a bit is spread only over the $L$ nonzero chips. In general, $L$ is much smaller than $N$ and thus the spreading is sparse. When $L = N$, the spreading becomes the ordinary dense random spreading.

The large sparse-sequence (LSS) CDMA is obtained by letting $K$ and $N$ tending to infinity and their ratio kept a constant $\alpha = K/N \in (0, \infty)$. The amplitudes $A_k$ are fixed regardless of sequence selection. The channel load equals $\alpha =$

$K/(WT_b) = K/N$ bits/s/Hz. In [9]-[11], the BP algorithm is applied to detect **b**. The LSS-CDMA is impractical since the user number $K$ and the spectral spreading factor $N$ (or bandwidth $W$) are practically fairly small.

*B. QLSS-CDMA*

Now we propose to construct a quasi-large sparse-sequence (QLSS) CDMA by a scheme of bit extending and multiplexing, similar to that in multicode CDMA. Consider a CDMA system where $K$ and $N$ are *finite* and *fixed*. Bit period $T_b$ and signal amplitudes are also fixed. User $k$ multiplexes and simultaneously transmits $B_k$ bits $b_{kj}$, $j = 1, …, B_k$, which are extended by a factor of $B$ to occupy $B$ bit periods of $BT_b$ seconds and spread by the $BN$-chip sparse sequences $\mathbf{s}_{kj} \in \{-1/\sqrt{L}, 0, 1/\sqrt{L}\}^{BN}$ that has $L$ nonzero chips randomly located and equiprobably taking on $\pm 1/\sqrt{L}$'s. The signal amplitude for $b_{kj}$ is $A_{kj}$. During an extended bit period $BT_b$, the chip MF outputs a $BN$-dimensional vector

$$\mathbf{r} = \sum_{k=1}^{K}\sum_{j=1}^{B_k} \mathbf{s}_{kj} A_{kj} b_{kj} + \mathbf{m} \qquad (2)$$

where $\mathbf{m} \sim N(\mathbf{0}, \sigma^2 \mathbf{I}_{BN})$. The signal mode (2) can be also written in the form $\mathbf{r} = \mathbf{SAb} + \mathbf{m}$ with $\mathbf{S}$, $\mathbf{A}$ properly written.

In the special instance with $L = N$, the QLSS-CDMA becomes the QLRS-CDMA [6].

The bit multiplexing factors $B_k$ and the temporal spreading factor $B$ tend to infinity and their ratios $\beta_k = B_k/B$, $k = 1, …, K$, are fixed. Then the channel load is fixed as

$$\alpha = \frac{1}{WBT_b}\sum_{k=1}^{K} B_k = \frac{1}{N}\sum_{k=1}^{K} \beta_k \qquad (3)$$

bits/s/Hz.

The constructed QLSS-CDMA is statistically identical to the LSS-CDMA [9]-[11] with channel load $\alpha$. However, QLSS-CDMA is practical because the total spreading factor, $BN$ equal to the temporal spreading factor times the spectral spreading factor, can be arbitrarily large since $B$ can be arbitrarily large. This is true even when spectrum is not spread with $N = 1$.

The advantage of the QLSS-CDMA is that as $B$ increases, it eventually possesses the LML characteristic so that the linear-complex LAS detectors can approach the NP-hard GML detection and approach the single-user bound in the high SNR regime.

*C. The LAS detectors*

Since the QLSS-CDMA signal model can also be written as (1), we present the LAS detectors below based on (1). The MF bank $\mathbf{S}$ outputs a $K$-dimensional vector

$$\mathbf{y} = \mathbf{S}^T\mathbf{r} = \mathbf{RAb} + \mathbf{n}. \qquad (4)$$

$\mathbf{R} = \mathbf{S}^T\mathbf{S}$ is the crosscorrelation matrix of spreading sequences and $\mathbf{n} = \mathbf{S}^T\mathbf{m} \sim N(\mathbf{0}, \sigma^2 \mathbf{R})$ is a colored Gaussian noise vector.

Given $\mathbf{b}(n)$, a LAS detector updates a number of bits and obtains a new $\mathbf{b}(n+1)$ until reaching a fixed point. The likelihood gradient evaluated at the current vector $\mathbf{b}(n)$ is $\mathbf{g}(n) = -\mathbf{Hb}(n) + \mathbf{Ay}$ where $\mathbf{H} = \mathbf{ARA}$. Suppose that $L(n) \subseteq \{1, …, K\}$ is the set of bits to be updated at step $n$. If the vector $\mathbf{b}(n+1)$ differs from $\mathbf{b}(n)$ by the bits whose indices are in $L_p(n)$ $\subseteq L(n)$, the likelihood gradient at the next step can be computationally efficiently updated by

$$\mathbf{g}(n+1) = \mathbf{g}(n) + 2\sum_{i \in L_p(n)} b_i(n)\mathbf{H}_i . \qquad (5)$$

The following generalized LAS detector defines the family of LAS detectors.

*LAS detector*: Given $L(n) \subseteq \{1, …, K\}$ for $n \geq 0$ and an initial vector $\mathbf{b}(0) \in \{-1, 1\}^K$. Bits are updated by

$$b_k(n+1) = \begin{cases} +1, & k \in L(n), b_k(n) = -1, g_k(n) > t_k(n), \\ -1, & k \in L(n), b_k(n) = +1, g_k(n) < -t_k(n), \\ b_k(n), & \text{otherwise,} \end{cases} \qquad (6)$$

where the $k$th threshold at step $n$ is

$$t_k(n) = \sum_{j \in L(n)} |H_{kj}|, \qquad \text{for } k \in L(n) \qquad (7)$$

and $\mathbf{g}(n+1)$ is updated by (5) in which $L_p(n)$ is the index set of flipped bits in (6). $\mathbf{b}^*$ is the fixed point and is also the finally decided vector if $\mathbf{b}(n) = \mathbf{b}^*$ for all $n \geq n^*$ with some $n^* \geq 0$. □

A particular LAS detector is obtained by specifying $L(n)$, $n \geq 0$. A sequential LAS (SLAS) detector updates one bit in each step $|L(n)| = 1$ for all $n \geq 0$ and the $k$th threshold from (7) is $t_k = A_k^2$. A wide-sense sequential LAS (WSLAS) detector is obtained if $|L(n)| = 1$ for all $n \geq n'$ with some $n' \geq 0$. The SLAS as well as WSLAS detectors all are LML detectors with neighborhood size one. Many other special LAS detectors can also be specified.

An upper bound on the BER of any LAS detector is obtained in [6] and the achievability of the optimum performance by the WSLAS detectors in the QLRS-CDMA is analyzed in [8]. It is obtained that the QLRS-CDMA possesses the LML characteristic. Furthermore, if $\alpha < \frac{1}{2} - 1/(4\ln 2)$ and $\sigma^2$ is vanishing at the same rate of $B$ increasing, an LML is almost surely the GML. The asymptotic multiuser efficiency (AME) of the WSLAS detectors converges almost surely to one. Simulation results verify the analytical results in a much broader region of $\alpha \leq 1.0$ and of any SNR with 500 bits. The per-bit complexity in all simulations is less than 0.5 times bit number.

Since the QLRS-CDMA is a special instance of the QLSS-CDMA with $L = N$, we conjecture that the QLSS-CDMA also possesses the LML characteristic so that the optimum performance and single-user bound are achievable by the WSLAS detectors. Moreover, the complexity of LAS detectors is mainly due to the core operation (5). The sequence sparsity in the QLSS-CDMA makes most elements of **H** zeros and therefore shall significantly reduce the complexity.

### III. SIMULATION RESULTS

In the simulation, we consider $B_k = B$ for all users. The SLAS detector with cyclical bit update is employed. The initial detector is the MF. The BER of the fixed point for each of the LAS detectors is reported. Users have equal power and then the BER is averaged over all users. For $BK \leq 128$, a set of sparse sequences is randomly selected in each transmission. For $BK > 128$, five sets of sparse sequences are randomly selected and fixed for all transmissions, and the BER's are estimated and shown together with their averages. As a complexity measure, the number of additions per bit counted from the core operation (5) is also estimated. In all simulations, only the multiplication of $BK$ is given and thus the results are applicable to any pair of integers $B$ and $K$ with their multiplication equal to the given $BK$. As a reference, the GML BER with the dense spreading, which is obtained by the statistic mechanics approach [3][4], is also shown.

Fig. 1 (a), (b) demonstrate respectively the BER and complexity versus bit number $BK$ with $\alpha = 0.8$ bits/s/Hz, SNR = 11 dB. As $BK$ increases, the BER's for all $L$ monotonically decrease. This justifies our proposal to construct QLSS-CDMA. When $L$ increases, the BER approaches the limit of the dense spreading with $L = N$, which approaches the GML BER and the single-use performance in high SNR when $BK \geq 500$. However, when $L = 16$, the BER is already very close to the GML BER with dense spreading. The complexity (additions/bit) monotonically increases with increasing $BK$ and $L$ but is saturated with respect to $BK$ for small $L$ ($\leq 16$). Hence, using sparse sequences with $L = 16$ can approach the GML BER while the complexity is significantly reduced in comparison with the dense spreading. The same result is demonstrated in Fig. 2 where (a), (b) respectively show the BER and complexity versus the nonzero-chip number $L$ with $\alpha = 0.8$ and SNR = 11dB.

Fig. 3 shows the BER versus SNR with $BK = 1024$ and $\alpha = 0.8$. As $L$ increases, the BER monotonically decreases. In particular, the SLAS BER with $L = 16$ is already indistinguishable from the GML BER with the dense spreading and approaches the single-user bound in the high SNR regime. The complexity of the SLAS detector is insensitive to SNR and the value can be seen from Figs. 1 and 2.

Since BER's are given with a fixed $BK$ without specifying $B$ and $K$, the result can be interpreted differently for different $K$. In all the simulations, the WSLAS detectors approach almost the same performance of the SLAS detector since both are LML detectors. However, the former has a little lower complexity. All the simulations show that the variation of BER's over different sequence samples is large when $L$ is small (say $L = 4, 6$) but decreases as $L$ increases.

## IV. CONCLUSIONS

We have recently proposed the quasi-large random sequence (QLRS) CDMA for the linear-complex LAS detectors to approach single-user performance with a practical complexity. The QLRS-CDMA employs the sequences that spread both in frequency and time. The QLRS-CDMA can be implemented in practice even when there is no spreading in spectrum with spectral spreading factor equal to one. It is interesting that the LAS detectors as well as the LML detectors can exploit the LML characteristic of QLRS-CDMA to approach the GML BER as well as the single-user bound in the high SNR regime. It is practically promising that the LAS detectors can be implemented with a linear complexity.

We further propose the quasi-large sparse-sequence (QLSS) CDMA by replacing random dense sequences in QLRS-CDMA with random sparse sequences. Simulation results show that the BER decreases monotonically and complexity increases as the number of nonzero chips $L$ increases. When $L = 16$, the BER is already very close to that of the QLRS-CDMA, approaching the single-user bound in the high SNR regime; and the complexity is significantly reduced by several orders due to sequence sparsity. The proposed QLSS-CDMA is promising to practical systems.

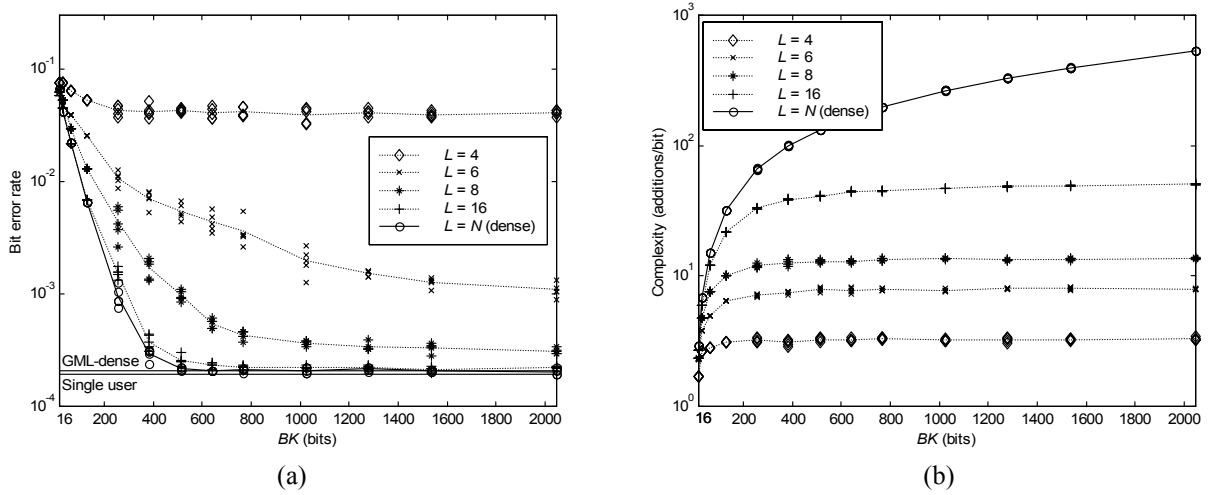

Fig. 1. (a) BER and (b) complexity versus bit number $BK$ with channel load $\alpha = 0.8$ bits/s/Hz and SNR = 11 dB.

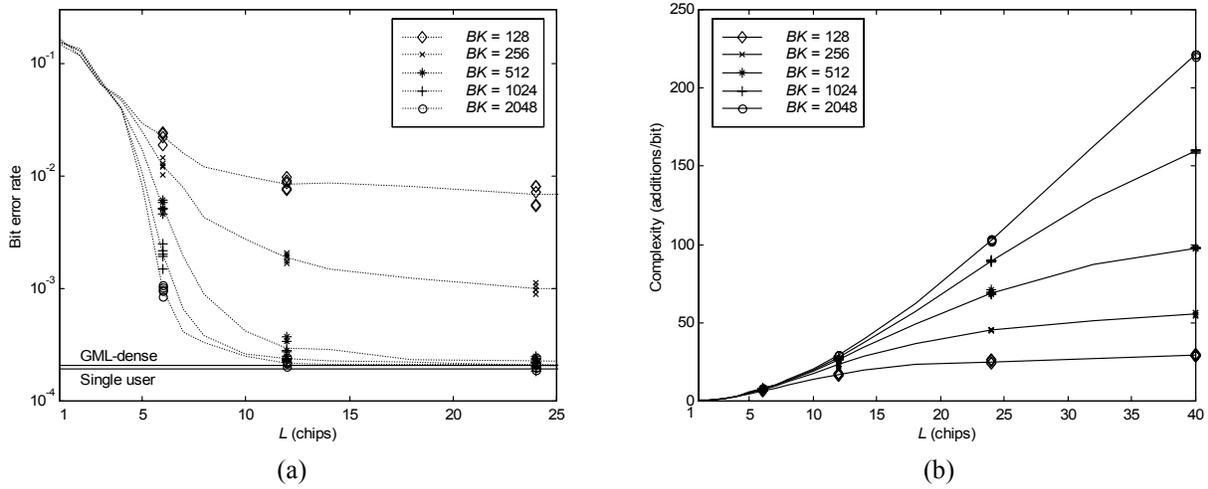

Fig. 2. (a) BER and (b) complexity versus nonzero-chip number $L$ with $\alpha = 0.8$ bits/s/Hz and SNR = 11 dB.

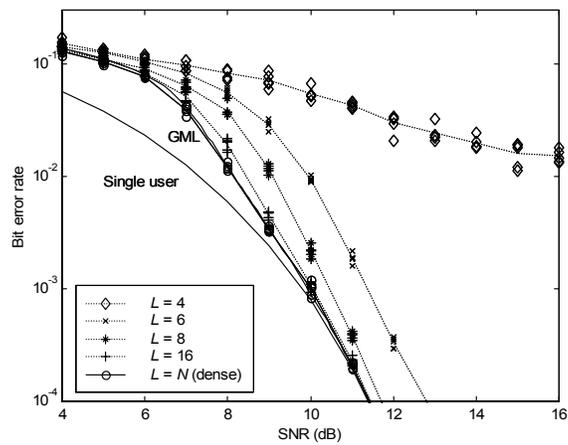

Fig. 3. BER versus SNR with $\alpha = 0.8$ bits/s/Hz and $BK = 1024$ bits.